\documentclass[12pt]{article}
\usepackage{graphicx}
\usepackage{cite}
\newcommand{\ep}{\epsilon}

\newcommand{\al}{\alpha_{s}}

\makeatletter

\@addtoreset{equation}{section}
\makeatother
\begin{document}
\setlength{\baselineskip}{16pt}
\begin{titlepage}
\begin{flushright}
HU-EP-10/36\\
SFB/CPP-10-61
\end{flushright}

\vspace{1.0cm}

\centerline{{\LARGE\bf Super AutoDipole}}

\vspace{25mm}

{\Large
\centerline{ K. Hasegawa
\footnote{e-mail : kouhei.hasegawa@physik.hu-berlin.de}}}
\vspace{1cm}
\centerline{{\it Institut f\"ur Physik, Humboldt-Universit\"at zu Berlin,}}
\centerline{\it D-10099 Berlin, Germany}
%
%

%
%   Abstract
%
\vspace{2cm}
\centerline{\large\bf Abstract}
\vspace{0.5cm}
The publicly available package for an automated dipole subtraction, 
AutoDipole, is extended to include the SUSY dipoles in the MSSM. 
All fields in the SM and the MSSM are available.
The code is checked against the analytical expressions for a simple process. 
The extended package makes it possible to compute the QCD NLO corrections to SUSY 
multi-parton processes like the stop pair production plus jets at the LHC. 
 
\end{titlepage}

%
% Body
%
%%%%%%%%%%%%%%%%%%%%%%
\section{Introduction \label{s1}}
%%%%%%%%%%%%%%%%%%%%%% 

The hierarchy problem of the standard model (SM) can be solved by supersymmetry (SUSY). 
Its minimal realization is the minimal supersymmetric standard model (MSSM)
and comprehensive reviews are given 
in~\cite{Haber:1984rc,Martin:1997ns,Weinberg:2000cr}.
In order for the hierarchy problem to be solved naturally  
the SUSY breaking scale and the masses of the super-partners are expected to 
be roughly 1 TeV at most. If this expectation is true,
the on-going LHC experiment is able to produce the super-partners directly and
discover them. What is required on the theoretical side is to make
both predictions of the various SUSY signals and the SM backgrounds
as precise as possible. The QCD NLO corrections are generally mandatory 
for the reliable predictions.

At the modern colliders like the LHC and the future ILC the automatization
of theoretical prediction is essentially required.
One reason is that the number of modes are too large to be treated manually
case by case.
The other reason is that the QCD NLO corrections involve more ingredients
than the LO case and especially the corrections to the multi-parton
processes are too lengthy and too complex to be treated by hand.
The automatization into a computer code can be realized only by general and
practical procedures which are executed in some simple algorithmic steps.
Here we briefly review the publicly available codes for the fixed order 
perturbative QCD. The LO (tree level) has already been automated well 
in Refs.~\cite{Stelzer:1994ta,Maltoni:2002qb,Alwall:2007st,
Hahn:1998yk,Hahn:2000kx,
Pukhov:1999gg,
Kanaki:2000ey,Yuasa:1999rg,
Moretti:2001zz,Kilian:2007gr,
Krauss:2001iv,Gleisberg:2008ta,
Mangano:2002ea}.
Some of these references include also the automation of the phase space integration,
the parton shower, the decay, and so on, i.e. they are event generators.
The NLO real emission corrections are also obtained by the same packages
because the real corrections enter at tree level.
About the NLO virtual 1-loop corrections, some essential ingredients are automatized and
the automation is now approaching perfection~\cite{Hahn:1998yk,Hahn:2000kx,
Yuasa:1999rg,Berger:2008sj,Binoth:2008uq,vanHameren:2009dr}.
Each of the real and virtual corrections possesses soft and collinear singularities
and they cancel each other for observables
~\cite{Bloch:1937pw,Kinoshita:1962ur,Lee:1964is}.
For the multi-parton processes like $2 \to 4$ partons, it is almost impossible
to realize the cancelation in a fully analytical way. Furthermore the inclusion of 
a jet algorithm makes the analytical phase space integration hopeless.
Therefore a general and practical procedure to treat such soft/collinear singularities
is desirable and was discovered in Refs.~\cite{Catani:1996vz,Catani:2002hc,
Phaf:2001gc,Frixione:1995ms,Frixione:1997np,
Giele:1991vf,Giele:1993dj,Keller:1998tf,Harris:2001sx}.
Among these references especially the Catani-Seymoure dipole subtraction~\cite{Catani:1996vz,
Catani:2002hc} is widely used for NLO predictions of the multi-parton 
processes in the so called the `NLO Wishlist' for LHC~\cite{Bern:2008ef}.
During the last two years the NLO corrections to the $2 \to 4$ processes
in the wishlist have been successfully obtained using the dipole subtraction
~\cite{Bredenstein:2008zb,Bredenstein:2009aj,
Berger:2009zg,Berger:2009ep,Berger:2010vm,
Ellis:2009zw,
Bevilacqua:2009zn,Bevilacqua:2010ve,
Binoth:2009rv}.
In order to apply the procedure to the various multi-parton processes, 
automation is required due to the increasing number of the subtraction terms. 
The algorithm and the formulas are sufficiently simple to implement into a computer code.
We have automated the procedure in the package, AutoDipole, and made it publicly 
available~\cite{Hasegawa:2008ae,Hasegawa:2009tx}. The source code is written in 
Mathematica and to obtain the reduced Born matrix elements, 
the package interfaces with 
the stand-alone version of the MadGraph~\cite{Stelzer:1994ta,Alwall:2007st}.
There are the other four packages to automate the subtraction procedure in the frameworks of 
SHERPA~\cite{Gleisberg:2007md},
TeVJet~\cite{Seymour:2008mu},
MadGraph/Event~\cite{Frederix:2008hu,Frederix:2010cj},
and HELAC~\cite{Czakon:2009ss}.
The FKS subtraction~\cite{Frixione:1995ms,Frixione:1997np} is also 
automatized in the framework of MadGraph/Event~\cite{Frederix:2009yq}.

Next let us consider the status of the automatic calculation for
the SUSY processes, typically in the MSSM. 
The number of fields and interactions in the MSSM are more
than double in comparison with the SM case, and the number of production
and decay modes are also much increased. Then the automatic calculation 
is really needed.
It is not simple even to calculate the spectrum of the SUSY masses and couplings
by hand and automation is realized~\cite{Baer:1999sp,Allanach:2001kg,
Porod:2003um,Djouadi:2002ze}.
The LO and the real corrections are well automatized as in the SM 
case~\cite{Beenakker:1996ed,Djouadi:1997yw,Heinemeyer:1998yj,
Lee:2003nta,Muhlleitner:2003vg,Belanger:2004yn,Gondolo:2004sc,
Ellwanger:2004xm,Mahmoudi:2007vz,Degrassi:2007kj}.
Event generators also have been 
constructed~\cite{Alwall:2007st,Yuasa:1999rg,
Mrenna:1996hu,Katsanevas:1997fb,
Hahn:2001rv,Corcella:2000bw,Moretti:2002eu,
Boos:2004kh,Reuter:2005us,Hagiwara:2005wg,Cho:2006sx,
Sjostrand:2006za}.
About the 1-loop virtual correction, the automation has been proceeded by
the packages for some restricted processes in the colliders~\cite{Beenakker:1996ed},
for the general processes~\cite{Hahn:2001rv}, and for the decay processes
~\cite{Heinemeyer:1998yj,Muhlleitner:2003vg,Degrassi:2002fi,
Frank:2006y,Iizuka:2010bh}.
About the general treatment of the soft/collinear singularities in 
the QCD NLO corrections of the SUSY processes,
the extension of the dipole subtraction is constructed~\cite{Catani:2002hc}.
However, there is no available implementation of the SUSY cases so far.
The purpose of this article is to provide the first publicly available package
of the dipole subtraction including SUSY processes in the MSSM by extending the 
AutoDipole package.
All dipole terms in the procedure include the reduced Born matrix elements and 
AutoDipole interfaces with MadGraph to obtain the matrix elements.
Since MadGraph includes the MSSM~\cite{Cho:2006sx}, the SUSY extension of 
AutoDipole should be straightforward and we realize it as SuperAutoDipole
package in the present article. 
The article is organized in the following way: In Section 2, the 
extension of the dipole subtraction into the MSSM is briefly reviewed
with a simple process. In Section 3, the usage of SuperAutoDipole is
explained and some benchmark numbers are shown. 
Section 4 is devoted to the conclusion.

%%%%%%%%%%%%%%%%%%%%%%
\section{Dipole subtraction in the MSSM \label{s2}}
%%%%%%%%%%%%%%%%%%%%%%

Let us start this section with the quick review of the MSSM.
The gauge groups are the same as in the SM. The gauge bosons (spin $s=1$)
are uniquely embedded in the vector multiplets and the super-partners
are the gauginos ($s=1/2$). The quarks and leptons ($s=1/2$) are included in
the chiral multiplets and the partners are the squarks and 
sleptons ($s=0$), respectively. The higgs boson ($s=0$)
is included in the chiral multiplet and the partner is
a higgsino ($s=1/2$). The sector of Yukawa interaction in the SM, 
which is the source of the fermion masses, is extended to
the SUSY version by introducing a superpotential.
In order to provide masses to both of the up- and down-type
quarks, a second higgs doublet is required.
To avoid baryon and lepton number violation, a symmetry called 
R-parity is imposed, which forbids the proton decay.
A small amount of R-parity violation is still allowed in
the parameter region where the prediction does not contradict 
with the data of the present experiments, as reviewed in 
Ref.~\cite{Dreiner:1997uz,Dreiner:2010ye}.
Hereafter we think only about the MSSM with exact R-parity.
Furthermore, explicit soft SUSY breaking terms are introduces, 
typically adding the mass terms of the super-partners.
The masses of the super-partners are assumed to be 
roughly at the TeV scale, which is the proper scale to solve the
hierarchy problem.
\begin{table}[h]
  \begin{center}
\begin{tabular}{|c|cccc|c|}
      \hline
(field, super-partner )& \mbox{spin:} & 1   & 1/2           & 0           &  $SU(3)_{c}, SU(2)_{L}$ \\ \hline
(quark, squark)        & &     &  ($q_{L}$,          & $\tilde{q}_{L})$ & 3, 2  \\
                       & &     &  ($q_{R}$,          & $\tilde{q}_{R})$ & 3, 1  \\ \hline
(gluon, gluino)        & & ($g$, &  $\tilde{g}$)  &             & 8, 1  \\
      \hline
\end{tabular}
    \caption{ The spins and the representations in the gauge group $SU(3)_{c}$ and $SU(2)_{L}$ are shown.
              The left-handed quark $q_{L}$ and the super-partner $\tilde{q}_{L}$
              represent $SU(2)_{L}$-doublet as $q_{L}=(u,d)_{L}$ and $\tilde{q}_{L}=(\tilde{u},\tilde{d})_{L}$, 
              respectively. The right-handed quark $q_{R}$ and the partner $\tilde{q}_{R}$ are 
              $SU(2)_{L}$-singlet as $q_{R}=u_{R}/d_{R}$ and $\tilde{q}_{R}=\tilde{u}_{R}/\tilde{d}_{R}$.
              The quarks and squarks have the copy of two more generations as $(c,s)$ and $(t,b)$.
\label{tablecolor}}
  \end{center}
\end{table}
We here take a closer look at the colored super-partners.
The super-partners of the quark and gluon are squark and
gluino, respectively, and they are summarized in Table~\ref{tablecolor}.
The masses are roughly at the TeV scale as written above.
The fields which possess same quantum numbers can be mixed. 
The mixings between the squarks $\tilde{q}_{L}$ and $\tilde{q}_{R}$ are 
large in the case of stop and sbottom.
The mass eigenstates are written as $\tilde{t}_{1(2)}$ and $\tilde{b}_{1(2)}$,
respectively. The mixing in the first and second generation can be
omitted. The gauge interactions of the squark and gluino 
are derived from the corresponding covariant derivatives, 
$|D_{\mu}\tilde{q}|^2$ and $\bar{\tilde{g}} D \!\!\!\! / \ \tilde{g}$, respectively.
The interactions induce the gluon emission from the squark and
gluino legs. The gluon emission causes the soft/collinear singularity
in the real emission correction.
The appearance of the new source of the singularity in the MSSM requires
new dipole terms in the dipole subtraction.

\begin{figure}[htbp]
  \begin{center}
    \leavevmode
     \includegraphics[width=0.79\textwidth]{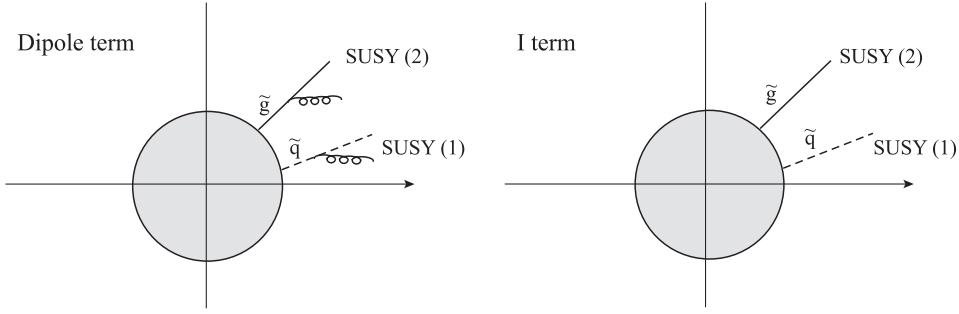}
    \caption{\small
      The emitter of the dipole terms (I-terms) for the SUSY processes are shown
      in the left (right). The symbols $\tilde{{\tt q}}$ and $\tilde{{\tt g}}$ 
      represent the squark and gluino respectively.}
    \label{all_susy}
  \end{center}
\end{figure}

The dipole terms for the massive quarks in the SM and the ones for the squarks and gluino 
typically in the MSSM were constructed at the same time~\cite{Catani:2002hc}.
As the squarks have mass but no spin, the dipole term is a bit simpler than
the massive quark case and the color factor is same as the one of the quark.
As the gluino has mass and the spin of one half, the 
kinematical part of the dipole is same as the massive quark case
and the color factor is same as the one of the gluon.
Since the spins of the colored super-partners are zero or one half 
but not one, the SUSY dipoles do not
have the correlation of the emitter helicity between the amplitude
and the conjugate. All explicit formulas for the SUSY dipoles are given 
in~\cite{Catani:2002hc}.
The algorithm to create the SUSY dipoles is also straightforward extension
from the SM case. Furthermore, since the squarks and gluino are sufficiently heavier
than the top quark in the MSSM, we do not have to take care of the massless limit
in practice. Then all dipoles for the SUSY cases are of the same type as the gluon emission, which is
sometimes called the diagonal splitting type. In the AutoDipole package we
categorize such diagonal splitting type of dipoles as `dipole 1'~\cite{Hasegawa:2009tx}.
Following the category we add two kinds of the SUSY dipoles to the `dipole 1' as shown 
at the left in Fig.~\ref{all_susy}.
For example, the dipole term with the gluon($j$) emission from the emitter squark($i$)
in the final state is the case of SUSY (1) at the left in Fig.~\ref{all_susy}. 
If the dipole has the spectator($k$) in the final state, 
it is written as $\mbox{D}_{ij,k}=\mbox{D}_{squark \ gluon,k}$ 
and the explicit form is written as
\begin{eqnarray}
\mbox{D}_{ij,k} = - \ \frac{1}{s_{ij}} \ \mbox{V}_{ij,k} \ \mbox{B1}(ij,k) \label{dipform}
\end{eqnarray}
with the dipole splitting function,
\begin{eqnarray}
\mbox{V}_{ij,k} = 8 \pi \mu^{2 \ep} \al 
\biggl[ 
 \frac{2}{1 - z_{i} (1 - y_{ij,k})} -
 \frac{\tilde{v}_{ij,k}}{v_{ij,k}}
\biggl( 
 2 + \frac{2 m_{i}^{2}}{s_{ij}}
\biggr)
\biggr],
\end{eqnarray}
where $s_{ij}=2p_{i} \cdot p_{j}$ and the quantities, $z,y,v,\tilde{v}$, are 
defined in~\cite{Catani:2002hc}. 
The $\mbox{B1}(ij,k)$ is the color linked Born squared (CLBS) and is defined as
\begin{eqnarray}
\mbox{B1}(ij,k) \ = \ \langle \mbox{B1} \ |\mbox{T}_{ij} \cdot \mbox{T}_{k} | \ \mbox{B1} \rangle,
\end{eqnarray}
where the $|\mbox{B1} \rangle$ represents the amplitude of the reduced Born process in the 
category `dipole1', which is made by the removal of one gluon in the final state
from the given real emission process.
The color operator $\mbox{T}_{ij}(\mbox{T}_{k})$ acts on the reduced Born amplitude 
$|\mbox{B1}\rangle$ and inserts the color factor into the emitter(spectator) line.
The Casimir operators in the dipole terms always cancel in the `dipole1' and
they are eliminated in Eq. (\ref{dipform}) for simplicity.

The 1-loop virtual correction owns soft and collinear 
singularities which originate from the loop integration and 
they are regulated as the poles $1/\ep$ and $1/\ep^2$
in the dimensional regularization. The dipole subtraction
procedure provides the ${\mbox{\bf I}}$-terms to subtract the poles
from 1-loop squared matrix element.
Corresponding to the SUSY dipole terms two kinds of ${\mbox{\bf I}}$-terms
are added to the SM ones as shown at the right in Fig.~\ref{all_susy}.
For example the ${\mbox{\bf I}}$-term with the emitter of the squark(i) and
the spectator of a massless parton(k),
${\mbox{\bf I}}_{i,k}={\mbox{\bf I}}_{squark, k}$, belongs to SUSY(1) category.
The ${\mbox{\bf I}}$-term is explicitly written as  
\begin{eqnarray}
\mbox{\bf I}_{i,k} &=& -\frac{\al}{2\pi}\frac{(4\pi)^{\ep}}{\Gamma(1-\ep)}
\frac{1}{C_{F}} \biggl[
C_{F}\biggl(\frac{\mu^{2}}{s_{ik}}\biggr)^{\ep}\biggl({\cal V}_{i,k}(m_{i}) - \frac{\pi^2}{3}\biggr)+  
\nonumber \\
& & \ \ \ \ \ \ \ \ \ \ \ \ \ \  \Gamma_{\tilde{q}}(m_{i}) + \gamma_{\tilde{q}} \ln \frac{\mu^2}{s_{ik}} + 
\gamma_{\tilde{q}} + K_{\tilde{q}} \biggr]  \mbox{B1}(i,k) 
\end{eqnarray}
with ${\cal V}_{i,k}(m_{i})= {\cal V}S_{i,k}(m_{i}) + {\cal V}NS_{i,k}(m_{i})$.
Here the singular part ${\cal V}S$ includes the poles as
\begin{eqnarray}
{\cal V}S_{i,k}(m_{i}) &=& \frac{1}{2\ep^2} + \frac{1}{2\ep} \ln \frac{m_{i}^2}{s_{ik}}
- \frac{1}{4}\ln^2 \frac{m_{i}^2}{s_{ik}} - \frac{\pi^2}{12} \nonumber \\
& &-\frac{1}{2}\ln\frac{m_{i}^2}{s_{ik}}\ln\frac{s_{ik}}{Q_{ik}^2}
-\frac{1}{2}\ln\frac{m_{i}^2}{Q_{ik}^2}\ln\frac{s_{ik}}{Q_{ik}^2}.
\end{eqnarray}
The non-singular part ${\cal V}NS$, and the other quantities,
$\Gamma_{\tilde{q}}(m_{i})$, $\gamma_{\tilde{q}}$, $K_{\tilde{q}}$ are defined in 
Refs.~\cite{Catani:2002hc}.
The CLBSs in a dipole term and in the corresponding
${\mbox{\bf I}}$-term are the same function of the input momenta.
The difference is that the CLBS in the dipole term
has as input the reduced momenta from the phase space
of the real correction process and one in the ${\mbox{\bf I}}$-term
has as input the momenta of the LO phase space.
The (Super) AutoDipole interfaces with MadGraph to obtain 
CLBS in four dimension. Then the generated ${\mbox{\bf I}}$-term by the
package belongs to the 't Hooft-Veltman scheme(tHV) in
the dimensional regularization. The 1-loop correction
also must be calculated in the same scheme.
But the direct application of this scheme to SUSY process
causes a problem of SUSY breaking, because
in the scheme the gluon running in the loop is treated 
as D-dimensional but the super-partner, gluino is treated
as 4-dimensional. Fortunately the problem is solved
by introducing a counter term in such a way as to 
restore SUSY at the level of the physical amplitude
in Refs.~\cite{Martin:1993yx,Beenakker:1996ch}.
About the ${\mbox{\bf P}}$- and ${\mbox{\bf K}}$-terms their emitter is only a massless parton 
in the initial state and the squarks and gluino in the final state can be only their spectator. 
The kinematical part
of the ${\mbox{\bf P}}$-term is insensitive to the spin of the spectator and the formula of the SM case
is not changed for the SUSY extension. Then only the ${\mbox{\bf K}}$-term must be extended
in the case that the spectator is a squark or gluino. 
Again all formulas for the dipole, ${\mbox{\bf I}}$-, ${\mbox{\bf P}}$- and ${\mbox{\bf K}}$- terms 
are given in~\cite{Catani:2002hc}.

In the remainder of this section, to demonstrate the SUSY cases explicitly
we take one of the simplest example, the stop pair production, 
$e^{-}e^{+} \to \tilde{t}_{1} \tilde{t}_{1}^{*}$, 
and will show the analytical expressions of the squared matrix element of the
real correction $|\mbox{M}|_{\mbox{{\tiny real}}}^2$, 
the dipole term $\mbox{D}$, and the subtracted 
real cross section $\bar{\sigma}_{\mbox{{\tiny real}}}$. 
The squared matrix element of the real correction is calculated as
\begin{eqnarray}
|\mbox{M}(e^{-}(1)e^{+}(2) \to \tilde{t}_{1}(3) \tilde{t}_{1}^{*}(4)g(5))|^2 &=&
256\pi^3 \alpha_{e}^2 Q_{t}^2 \al N_{c}C_{F} \frac{1}{s_{12}^2 s_{35}^2 s_{45}^2} 
\cdot \mbox{A} \label{msq}
\end{eqnarray}
with
\begin{eqnarray}
\mbox{A} = s_{12} s_{35} s_{45} \bigl[
s_{12}(-s_{45} + 2s_{14} + s_{15}) + s_{35}( s_{45} - s_{14} - s_{15}) + 
s_{14}(s_{45} - 2s_{14} - 2s_{15}) \bigr]   \nonumber\\
 -2m^2\bigl[ 
s_{12}^2 s_{35} s_{45} +s_{12}(s_{35}^2(- s_{45} + s_{14} + s_{15}) + 
s_{35}s_{45}(-s_{45} + 2s_{14} + s_{15})
+ s_{45}^2s_{14})  \nonumber \\
 \ \ \ \ \ \ \ \ \ \  - s_{35}^2 (s_{14} + s_{15})^2 - 
2s_{35}s_{45}s_{14}(s_{14} + s_{15}) -s_{45}^2 s_{14}^2 \bigr]
+ 2m^4 s_{12}(s_{35}+s_{45})^2
\end{eqnarray}
which is uniquely expressed by the minimal set of the independent Lorentz
scalars $s_{ij}$ and the stop mass $m$. 
Here we take the five scalars ($s_{12},s_{14},s_{15},s_{35},s_{45}$) as the set.
The other quantities are written as 
the electromagnetic coupling constant $\alpha_{e}$,
the charge of the stop $Q_{t}=2/3$,
the strong coupling constant $\al$,
and the usual color factors $N_{c}=3$ and $C_{F}=4/3$.
One dipole for the real correction is calculated as
\begin{eqnarray}
\mbox{D}_{35,4}&=&-\frac{8\pi \al}{s_{35}} \Biggl[ \frac{2s_{345}}{s_{35}+s_{45}} - 
\frac{s_{345} - s_{35}}{s_{345}}
\biggl( \frac{2m^2}{s_{35}} + 2 \biggr) \nonumber\\
&& \hspace{30mm} \sqrt{\frac{s_{345}^2 - 4m^4}{(s_{345}-s_{35})^2  - 4m^2 s_{35} - 4m^4}} 
\Biggr] \ \mbox{B}1(35,4) \label{dip1}
\end{eqnarray}
with the CLBS,
\begin{eqnarray}
\mbox{B}1(35,4)&=& -C_{F} \cdot |\mbox{M}(e^{-}(1)e^{+}(2) \to 
\tilde{t}_{1}(\tilde{35}) \tilde{t}_{1}^{*}(\tilde{4}))|^2 \nonumber\\
&=& -C_{F} N_{c} 8\pi^2 \alpha_{e}^2 Q_{t}^2 \frac{1}{s_{12}^2} 
\biggl[s_{12}^2 - 4m^2 s_{12} \nonumber\\
&&- \frac{4(s_{345}^2 - 4m^4)}{(s_{345} + 2m^2)(s_{345} - 2s_{35} -2m^2)+s_{35}^2}
\biggl(s_{14} - \frac{(s_{345} - s_{35} + 2m^2)s_{12}}{2s_{345} + 4m^2}\biggr)^2
 \biggr], \nonumber\\
\label{b1}
\end{eqnarray}
where the reduced Born amplitude $\mbox{M}(e^{-}e^{+} \to \tilde{t}_{1}\tilde{t}_{1}^{*})$ 
has the reduced momenta $\tilde{p}_{35}$ and $\tilde{p}_{4}$
for the stop and anti-stop. The $s_{345}$ is defined as $s_{345}=s_{34}+s_{45}+s_{53}$.
One more dipole $\mbox{D}_{45,3}$ is obtained by the trivial replacement of 
$(3 \leftrightarrow 4)$ in Eq. (\ref{dip1}), and the replacements of
$(3 \leftrightarrow 4)$ and $(1 \leftrightarrow 2)$ in Eq. (\ref{b1}).
The collinear singularities of the squared matrix element in 
Eq. (\ref{msq}) are screened by the masses of the stop and anti-stop,
and the soft singularity is subtracted by two dipoles as in Eq. (\ref{dip1}).
Then the phase space integration of the subtracted quantity becomes finite. 
In order to see the finiteness concretely we here perform the phase space
integration in an analytical way. For simpler integration we introduce
the averaged lepton tensor for the initial electron-positron current and 
obtain the subtracted infrared (IR) safe real cross section as
\begin{eqnarray}
\bar{\sigma}(e^{-}e^{+} \to \tilde{t}_{1} \tilde{t}_{1}^{*}g)_{\mbox{{\tiny real}}}
&\equiv& \int d \Phi^{(3)} \ \biggl( \ |\mbox{M}(e^{-}e^{+} \to \tilde{t}_{1}\tilde{t}_{1}^{*}g)|^2
\ - \ \mbox{D}_{35,4} - \mbox{D}_{34,5} \biggr) \nonumber\\
&=& \sigma_{0} 
\frac{\al}{\pi} C_{F} \cdot \mbox{I} \label{anal}
\end{eqnarray}
where the measure of the phase space $d \Phi^{(3)}$ includes the other factors as the fluxes
and the LO cross section in the massless limit is written as 
$\sigma_{0}=\pi \alpha_{e}^2 Q_{t}^2 N_{c}/(3s_{12})$.
The mass- and energy-dependent part is calculated as 
\begin{eqnarray}
\mbox{I} &=& \frac{\sqrt{1-4\rho^2}}{1-2\rho^2}(1+4\rho-4\rho^2-16\rho^3+12\rho^4) \nonumber\\
&&-2(1-5\rho^2+\rho^4)\ln \frac{1-\sqrt{1-4\rho^2}}{2\rho} \nonumber\\
&&-2(1-4\rho^2)^{3/2} \ln \frac{1+2\rho}{\rho} \nonumber\\
&&+\rho^2(1-\rho^2) \ln \Biggl| \frac{1-3\rho^2-(1-\rho^2)\sqrt{1-4\rho^2}}{2\rho^3} \Biggr|
\end{eqnarray}
with $\rho=m/\sqrt{s_{12}}$. Then we can confirm the finiteness of the subtracted cross 
section with help of the analytical expression. In this way the soft/collinear singularities for
any real emission process in the SM and the MSSM are subtracted by the corresponding
dipole terms and the subtracted cross section becomes finite.
It is noted that for the multi-parton processes 
the analytical phase space integration
with a jet algorithm is almost impossible and the integration can be
performed generally by the Monte Carlo integration.
The AutoDipole package provides the Fortran routines of the
integrand, which can be directly integrated by the Monte Carlo method.

%%%%%%%%%%%%%%%%%%%%%%
\section{Usage of SuperAutoDipole \label{s3}}
%%%%%%%%%%%%%%%%%%%%%%
The extension of the algorithm for the SUSY cases is straightforward and all 
the formulas are given as seen briefly in the previous section.
The calculations of all ${\mbox{\bf D}}$-, ${\mbox{\bf I}}$-, ${\mbox{\bf P}}$- and 
${\mbox{\bf K}}$- terms in the MSSM 
are reduced to the tree level processes in the same model
as shown in Eq. (\ref{dipform}). 
In order to obtain the square of the reduced matrix 
element, our package interfaces with MadGraph. MadGraph automatically 
provides codes of the squared matrix elements for the general processes at tree level 
in the SM~\cite{Stelzer:1994ta}.
Later it is extended to include the processes in MSSM~\cite{Cho:2006sx} 
and some models beyond the SM~\cite{Alwall:2007st}. 
MadGraph generates all diagrams and writes down the Fortran routines to call 
the corresponding wave functions, vertices, and so on in the HELAS library
~\cite{Hagiwara:1990dw,Murayama:1992gi,Hagiwara:2008jb}.
The choice of the available models is well structured and
determined at the input file.
This feature of MadGraph allows the straightforward extension of the AutoDipole 
package to include the SUSY cases 
without changing the basic structure of the original one for the SM.
In the present section we would explain the usage step by step
with the example in the previous section focusing on the new commands.
The installation of SuperAutoDipole is exactly same as with AutoDipole.
The usage is also basically same as
\begin{verbatim}
1. Inclusion of driver : << driver_user.m
2. Setup of parameters : parameter.m
                         interactions.dat
                         param_card.dat
3. Run of package      : GenerateAll[{initial},{final}]
4. Run of the generated codes : make runD, runI, runPK, checkIR
\end{verbatim}
The steps 1. and 4. are exactly same and here we would explain mainly 
the new points at steps 2. and 3. \\

\begin{enumerate}
\item[] {{\bf Step 2. Setup of parameters}}
\vspace{5mm}
\\
{\tt parameter.m} : \\
   One new parameter is {\tt model} whose value can be {\tt sm} or {\tt mssm}.
   If the user input is a process in the MSSM, it must be set as {\tt model=mssm}.
   The default is {\tt model=sm} for the processes in the SM.
   One more new parameter is {\tt massiveb} which specifies the masssive({\tt 1}) or 
   massless({\tt 0}) of the bottom quark. Default is the massive case.
\vspace{5mm}
\\
{\tt interactions.dat} :\\ 
   The file is one of the MadGraph which is located at {\tt ./patch/Models/sm}
   ({\tt ./patch/ Models/mssm}) for the SM(MSSM) case. User can switch on/off the interactions
   in the SM/MSSM, which is the exactly same way with the use of the MadGraph. 
   For the example process we would switch off the Z propagation
   for simplicity. It is realized by the commenting out one line as \ 
   `{\tt \#e-  e-  z GZL QED}'.
\vspace{5mm}
\\
{\tt param\_card.dat} : \\
   The file is also in MadGraph which is at the same place with {\tt interactions.dat}.
   It specifies the coupling constants, the masses, 
   the decay widths, and so on.  
   If the parameter {\tt massiveb} at {\tt parameter.m} is set at massless({\tt 0}),
   the bottom quark mass at this file also should be set at zero.  
   The decay width of a participating heavy particle sometime breaks 
   the cancelation of soft/collinear singularities. Then in the example,
   the decay width of the stop should be switched off as \
   `{\tt DECAY   1000006     0.00000000E+00 \# stop1 decays}'.
   The parameters of the file can be changed after the generation of the code
   in each created directory under {\tt ./process}.
\vspace{5mm}
\item[] {{\bf Step 3. Run of SuperAutoDiplole}}\\
The package is run by the same command with the AutoDipole for the SM as
\begin{verbatim}
GenerateAll[{e,ebar},{st1,st1bar,g}],
\end{verbatim}
which generates the Fortran routines of all $\mbox{\bf D}$, $\mbox{\bf I}$, $\mbox{\bf P}$, and 
$\mbox{\bf K}$ terms.
The input process is the example one, $e^{-}e^{+} \to \tilde{t}_{1} \tilde{t}_{1}^{*}g$.
All fields in the SM and the MSSM are available and the input forms of 
the SM and MSSM  fields are summarized in 
Table~\ref{tablesm} and ~\ref{tablemssm} respectively in Appendix~\ref{a1}.
The input forms of the publicly available packages can be various, depending on
the author's convention, the programming language, and so on.
In order that the packages can be used by many users comfortably without confusion,
it is quite reasonable to unify the input form and the definitions of the usual parameters.
In the community of the automatic calculation of the MSSM processes, such necessity is
well understood and the efforts have been made
~\cite{Allanach:2002nj,Skands:2003cj,Allanach:2008qq}.
In the same direction the PDG numbering scheme for the SM, MSSM, and some 
popular extensions of the SM are defined in~\cite{Amsler:2008zzb}. 
Following the accord, our package also provides the input form
in the scheme by the command,
\begin{verbatim}
GenerateAllNumber[{11,-11},{1000006,-1000006,21}],
\end{verbatim}
which is exactly equivalent with the above command.
The list of all fields in the numbering scheme is also shown in 
Tables~\ref{tablesm} and~\ref{tablemssm}.
The command generates the Fortran code in the new directory
under {\tt ./process}. The name of the new directory follows the convention
of MadGraph, which is also shown in Tables~\ref{tablesm} and~\ref{tablemssm}. 
The evaluation of the generated code at the example process, 
$e^{-}e^{+} \to \tilde{t}_{1} \tilde{t}_{1}^{*}g$,
completely agrees with the analytical result in Eqs.~(\ref{msq}) and (\ref{dip1}).
In order to provide benchmarks for the SUSY dipoles,  
the values are shown on two phase space points in Table~\ref{tablecomp} in 
Appendix~\ref{a2}.
The package also provides the automatic check of the process on the user's machine
by the command,
\begin{verbatim}
./lib/sh/check_susy.sh.
\end{verbatim}
It confirms that the evaluation of the generated code on the user's machine 
agrees with the analytical results in Eqs. (\ref{msq}) and (\ref{dip1}) on five phase space points,
including two used in Appendix~\ref{a2}. When it is confirmed, the command returns the message,
\begin{verbatim}
|M|^2(Real) :Confirmed
 Dipole     :Confirmed.
\end{verbatim}
Some advanced users may prefer the run of the package from the shell.
It is possible by the shell script as
\begin{verbatim}
math << EOF
<<driver_user.m
GenerateAll[{e, ebar}, {st1, st1bar, g}]
Quit
EOF.
\end{verbatim}
Such a way is also used at the above mentioned shell script, {\tt check\_susy.sh}.
\end{enumerate}

The AutoDipole package provides all ${\mbox{\bf D}}$-, ${\mbox{\bf I}}$-, 
${\mbox{\bf P}}$- and ${\mbox{\bf K}}$- terms,
in other words it provides all integrands of the phase space integration,
except for the 1-loop virtual correction. 
The ${\mbox{\bf I}}$-terms are provided by the package in the tHV scheme
as mentioned in the Sec.~\ref{s2}. On the other hand there
is a more suitable scheme for the SUSY process, i.e. 
dimensional reduction (RD) where SUSY is preserved,
because the gluon and gluino are treated in same 
dimensionality. The transition rule from tHV to DR is known
in Refs. \cite{Kunszt:1993sd,Catani:2000ef,Signer:2008va}. 
The choice of a scheme by one input parameter
will be automated in the near future.
One advantage of the package 
is that the user can copy the created directory into an user's
directory and perform the Monte Carlo integration on his 
favorite setup to obtain the observables.
The feature easily allows also the user to
compare the results of his own code with the results of 
the generated code by the package.
Here in order to confirm the finiteness of the subtracted cross section
as well as the quality of the generated code,
we execute the VEGAS Monte Carlo integration of the example process
on a private setup.
The $5 \times 10^{8}$ phase space points times 10 iteration in
the VEGAS algorithm are evaluated with the double precision accuracy and 
the result is obtained in the unit of atto barn as
\begin{eqnarray}
\bar{\sigma}(e^{-}e^{+} \to \tilde{t}_{1} \tilde{t}_{1}^{*}g) = 
92.5719693 \pm 0.0001715 \ \mbox{[ab]} \label{nres}
\end{eqnarray}
where we use the default values of the MadGraph as written in the caption
of Table.~\ref{tablecomp} and the center-of-mass energy is set at 
$\sqrt{s_{12}}=4$[TeV].
The corresponding value of the analytical expression in Eq. (\ref{anal}) is
\begin{eqnarray}
\bar{\sigma} = 92.5719027 \ \mbox{[ab]} \label{ares}
\end{eqnarray}
The results in Eqs. (\ref{nres}) and (\ref{ares}) completely agree with the 
sufficient precision.
The agreement gives a guarantee about the quality of the generated codes.

Finally we would add the second benchmark numbers for the process, 
$g(1)g(2) \to \tilde{t}_{1}(3) \tilde{t}_{1}^{*}(4)$ $g(5)$ which
is a sub-partonic process of the real emission correction to 
the observable in a hadron collider, $pp \to \tilde{t}_{1} \tilde{t}_{1}^{*} + \mbox{X}$.
For this process 12 dipoles exist,
$\mbox{D}_{35,4},\mbox{D}_{35,1},\mbox{D}_{35,2},\mbox{D}_{45,3},\mbox{D}_{45,1},$ 
$\mbox{D}_{45,2},\mbox{D}_{15,3},\mbox{D}_{15,4},\mbox{D}_{15,2},$ 
$\mbox{D}_{25,3},\mbox{D}_{25,4},$ and $\mbox{D}_{25,1}$.
The values of the squared matrix $|\mbox{M}|^{2}$ and the sum of the dipoles $\sum \mbox{D}$ are
shown on two phase space points in Table~\ref{tablecomp} in Appendix~\ref{a2}. 
The command {\tt check\_susy.sh} includes also the check of these values on the user's machine.

%%%%%%%%%%%%%%%%%%%%%%
\section{Conclusion \label{s4}}
%%%%%%%%%%%%%%%%%%%%%%
We extend the AutoDipole package to include the SUSY dipoles for the processes
in the MSSM. 
The extended package, SuperAutoDipole, is the first publicly available one of 
automated dipole subtraction for the SUSY processes. 
All fields in the SM and the MSSM are available. 
The generated code for the stop pair production is checked against
the analytical expression including the phase space integration.
The package makes possible the QCD NLO corrections
to the SUSY multi-parton processes like $\tilde{t}_{1}\tilde{t}_{1}^{*}+ \mbox{jets}$
observables at the LHC. 
Thus the package can effectively help the detailed analysis of SUSY
at the LHC after they will be really discovered there.
The SuperAutoDipole package is available for download from~\cite{SuperAutoDipole:2010} or from 
the author upon request. The MadGraph package (stand-alone version) which includes the 
HELAS library may be obtained from~\cite{Madgraph:2009}.

\subsection*{\underline{Acknowledgments}}
I am grateful to S. Moch for useful discussions and careful reading of the article.
I am also indebted to U. Langenfeld for careful reading and helpful remarks.
This work is supported in part by the Deutsche Forschungsgemeinschaft in 
Sonderforschungs\-be\-reich/Transregio~9.

\newpage
\appendix
%
% -------------------------------------------------------------------
%
\section*{Appendix}
%
% -------------------------------------------------------------------
%

\section{Input form \label{a1}}
\begin{table}[h!]
\begin{tabular}{cccc}
Field & PDG & Input & Output \\
$d$ & 1 & d & d \\
$u$ & 2 & u & u \\
$s$ & 3 & s & s \\
$c$ & 4 & c & c \\
$b$ & 5 & b & b \\
$t$ & 6 & t & t \\
\\
$e^{-}$ & 11 & e & e-/e+ \\
$\nu_{e}$ & 12 & nue & ve \\
$\mu^{-}$ & 13 & muon & mu-/mu+ \\
$\nu_{\mu}$ & 14 & numu & vm \\
$\tau^{-}$ & 15 & tau & ta-/ta+ \\
$\nu_{\tau}$ & 16 & nutau & vt \\
\\
$g$ & 21 & g & g \\
$\gamma$ & 22 & gamma & a \\
$Z^{0}$ & 23 & Z & z \\
$W^{+}$ & 24 & Wp/Wm & w+/w- \\
\\
$h^{0}$ & 25 & h & h
\end{tabular}
    \caption{
The input forms for the SM fields are shown. 
The column of `PDG' shows the numbers in the PDG numbering scheme~\cite{Amsler:2008zzb}.
The numbers are used as the input form of the command, {\tt GenerateAllNumber[]}.
The number of the anti-particles is obtained by adding minus sign to the particle case.
For example, `{\tt -1}' is used for the anti-down quark.
The `Input' shows the input forms of the command, {\tt GenerateAll[]}.
The form of the anti-particle is obtained by adding `bar' to the last of the particle case.
For example, `{\tt dbar}' is for $\bar{d}$.
One exception is the case of $W^{\pm}$ and the forms are explicitly shown.
The `Output' shows the names of the created directly under {\tt ./process}.
The name for the anti-particles is obtained by adding `x' to the last.
For example, `{\tt dx}' is for $\bar{d}$.
The exceptions are $e^{-},\mu^{-},\tau^{-}$, and $W^{+}$, and shown explicitly.
   \label{tablesm}
      }
\end{table}

\newpage

\begin{table}[h!]
\begin{tabular}{cccc}
Field & PDG & Input & Output \\
$\tilde{d}_{L}$ & 1000001 & sdl & dl \\
$\tilde{u}_{L}$ & 1000002 & sul & ul \\
$\tilde{s}_{L}$ & 1000003 & ssl & sl \\
$\tilde{c}_{L}$ & 1000004 & scl & cl \\
$\tilde{b}_{1}$ & 1000005 & sb1 & b1 \\
$\tilde{t}_{1}$ & 1000006 & st1 & t1 \\
\\
$\tilde{e}_{L}$ & 1000011 & sel & el-/el+ \\
$\tilde{\nu}_{e L}$ & 1000012 & snuel & sve \\
$\tilde{\mu}_{L}$ & 1000013 & smul & mul-/mul+ \\
$\tilde{\nu}_{\mu L}$ & 1000014 & snumul & svm \\
$\tilde{\tau}_{1}$ & 1000015 & stau1 & ta1-/ta1+ \\
$\tilde{\nu}_{\tau L}$ & 1000016 & snutaul & svt \\
\\
$\tilde{d}_{R}$ & 2000001 & sdr & dr \\
$\tilde{u}_{R}$ & 2000002 & sur & ur \\
$\tilde{s}_{R}$ & 2000003 & ssr & sr \\
$\tilde{c}_{R}$ & 2000004 & scr & cr \\
$\tilde{b}_{2}$ & 2000005 & sb2 & b2 \\
$\tilde{t}_{2}$ & 2000006 & st2 & t2 \\
\\
$\tilde{e}_{R}$ & 2000011 & ser & er-/er+ \\
$\tilde{\mu}_{R}$ & 2000013 & smur & mur-/mur+ \\
$\tilde{\tau}_{2}$ & 2000015 & stau2 & ta2-/ta2+ \\
\\
$\tilde{g}$ & 1000021 & sg & go \\
\\
$\tilde{N}_{1}$ & 1000022 & sN1 & n1 \\
$\tilde{N}_{2}$ & 1000023 & sN2 & n2 \\
$\tilde{C}^{+}_{1}$ & 1000024 & sC1p/sC1m & x1+/x1- \\
$\tilde{N}_{3}$ & 1000025 & sN3 & n3 \\
$\tilde{N}_{4}$ & 1000035 & sN4 & n4 \\
$\tilde{C}^{+}_{2}$ & 1000037 & sC2p/sC2m & x2+/x2- \\
\\
$h^{0}$ & 25 & h & h1  \\
$H^{0}$ & 35 & H0 & h2  \\
$A^{0}$ & 36 & A0 & h3  \\
$H^{+}$ & 37 & Hp/Hm & h+/h-  
%$\tilde{G}$  & 1000039 & sgr &   
\end{tabular}
    \caption{
The input form for the MSSM fields are shown. 
The notation of the MSSM fields in the column of `Field' follows 
one in~\cite{Martin:1997ns}.
The definitions of the other three columns are same with the 
SM case in Table.~\ref{tablesm}. The exceptions are shown explicitly.
   \label{tablemssm}
      }
\end{table}

\section{Benchmark numbers \label{a2}}
The evaluations of the squared matrix elements and the dipole terms are shown in 
Table \ref{tablecomp} at two processes,
$e^{-}e^{+} \to \tilde{t}_{1} \tilde{t}_{1}^{*}g$ and $gg \to \tilde{t}_{1} \tilde{t}_{1}^{*}g$
on two phase space points, PSP1 and PSP2 in the unit of [GeV]. 
\begin{eqnarray}
\mbox{PSP1}: \ \ p_{1} &=& \scriptstyle (500,0,0,500),
\nonumber \\
p_2 &=& \scriptstyle (500,0,0,-500),
\nonumber\\
p_{3} &=& \scriptstyle
(465.4575514300951,88.15610495656038,197.5104870839467,-100.6674552035491)
\nonumber\\
p_{4} &=& \scriptstyle
(442.2757454441043,-9.535905415651449,-180.8891965873525,55.32717033368853)
\nonumber\\
p_{5} &=& \scriptstyle
(92.26670312580057,-78.62019954090894,-16.62129049659421,45.34028486986055)
\end{eqnarray}
\begin{eqnarray}
\mbox{PSP2}: \ \ p_{1} &=& \scriptstyle (500,0,0,500),
\nonumber \\
p_2 &=& \scriptstyle (500,0,0,-500),
\nonumber\\
p_{3} &=& \scriptstyle
(443.0578523569805,72.10954684770206,123.4569300473834,-126.9801398447153) 
\nonumber\\
p_{4} &=& \scriptstyle
(401.6307218084469,-0.27.27253369182125,-16.78202732992333,23.38599726779297)
\nonumber\\
p_{5} &=& \scriptstyle
(155.3114258345725,-44.83701315588080,-106.6749027174601,103.5941425769223)
\end{eqnarray}
\begin{table}[h!]
  \begin{center}
    \leavevmode
    \begin{tabular}[h]{|c|l|l|}
      \hline
                   &$|\mbox{M}|^{2} [\mbox{GeV}^{-2}]$ &$\sum \mbox{D} [\mbox{GeV}^{-2}]$ \\
      \hline\multicolumn{3}{|l|}
      {$e^{-}e^{+} \to \tilde{t}_{1} \tilde{t}_{1}^{*}g$}    \\
      %%----------------------------------------------
      %% PS 1
      \hline\multicolumn{3}{|l|}
      {PSP 1}  \\ \hline
      AutoDipole  & $3.38015568897016 \cdot 10^{-7}$ & $  4.38408930011336 \cdot 10^{-7}$ \\
      Analytics   & $3.38015568897014 \cdot 10^{-7}$ & $  4.38408930011337 \cdot 10^{-7}$ \\
      %%----------------------------------------------
      %%----------------------------------------------
      %% PS 2
      \hline\multicolumn{3}{|l|}
      {PSP 2} \\ \hline
      AutoDipole  & $7.46080394542307 \cdot 10^{-8}$ & $ -2.35778790281811 \cdot 10^{-7}$ \\
      Analytics   & $7.46080394542308 \cdot 10^{-8}$ & $ -2.35778790281812 \cdot 10^{-7}$ \\ \hline
      \multicolumn{3}{|l|}
      {$gg \to \tilde{t}_{1} \tilde{t}_{1}^{*}g$} \\ \hline
      %%----------------------------------------------
      %% PS 1
      PSP 1  & $2.83987176058353 \cdot 10^{-4}$ & $  2.77981806065659 \cdot 10^{-4}$ \\
      %%----------------------------------------------
      %%----------------------------------------------
      %% PS 2
      PSP 2 & $2.10737068989263 \cdot 10^{-4}$ & $ 2.01941234529644 \cdot 10^{-4}$ \\
      %%----------------------------------------------
      \hline
   \end{tabular}
    \caption{
      The results of the $|\mbox{M}(e^{-}e^{+} \to \tilde{t}_{1} \tilde{t}_{1}^{*}g)|^{2}$ 
      and the sum of two dipole terms $\sum \mbox{D}$ on PSP1 and PSP2 are shown.
      The first row is the evaluation of the automatically generated code by the package.
      The second row is the evaluation of the analytical expressions in Eqs. (\ref{msq}) and (\ref{dip1}).
      The results show the perfect agreement of 14 digits. 
      The values of the process $gg \to \tilde{t}_{1} \tilde{t}_{1}^{*}g$ are also shown at the last two entries. 
      We use the default values of the MadGraph as $\al= 0.118, \alpha_{e}=0.00781653039848672$ and 
      $m=399.668493$[GeV]. 
      These numbers can be reproduced on the user's machine by the command {\tt check\_susy.sh}.
       \label{tablecomp}
      }
  \end{center}
\end{table}
 
\newpage
 
{\footnotesize

%\bibliography{./lit_10}
%\bibliographystyle{./h-elsevier2}

}

\end{document}